\documentclass[prl,twocolumn,showpacs,preprintnumbers,amsmath,amssymb]{revtex4}

\usepackage{graphicx}
\usepackage{dcolumn}
\usepackage{bm}
\usepackage{color}

\begin{document}

\title{Fast Excitation and Photon Emission of a Single-Atom-Cavity System}

\author{J.~Bochmann}
\author{M.~M{\"{u}}cke}
\author{G.~Langfahl-Klabes}
\altaffiliation[Present address: ]
{Clarendon Laboratory, University of Oxford, Parks Road, Oxford OX1 3PU, UK}
\author{C.~Erbel}
\altaffiliation[Present address: ]
{Lehrstuhl f{\"{u}}r Energiesysteme, Technische Universit{\"{a}}t M{\"{u}}nchen, 85748 Garching, Germany}
\author{B.~Weber}
\author{H.~P.~Specht}
\author{D.~L.~Moehring}
 \email{david.moehring@mpq.mpg.de}
\author{G.~Rempe}

\affiliation{Max-Planck-Institut f{\"{u}}r Quantenoptik, Hans-Kopfermann-Strasse~1, 85748 Garching, Germany}

\date{\today}

\begin{abstract}
We report on the fast excitation of a single atom coupled to an optical cavity using laser pulses that are much shorter than all other relevant processes. The cavity frequency constitutes a control parameter that allows the creation of single photons in a superposition of two tunable frequencies. Each photon emitted from the cavity thus exhibits a pronounced amplitude modulation determined by the oscillatory energy exchange between the atom and the cavity. Our technique constitutes a versatile tool for future quantum networking experiments. 
\end{abstract}


\maketitle

Single atoms exchanging single optical photons are likely to be
the essential components for the processing of information in
distributed quantum networks \cite{zoller:2005}. Both carriers of
quantum information exhibit low decoherence rates and high
controllability for information stored, for example, in the spin
state of an atom and the polarization state of a photon. This
fact has made it possible to implement increasingly more complex quantum
protocols involving atom-photon entanglement \cite{blinov:2004,
volz:2006, moehring:2007b, matsukevich:2008, wilk:2007b}. These
experiments were performed in two different settings. One employed
single trapped ions or atoms in a free-space radiation environment
\cite{blinov:2004, volz:2006, moehring:2007b, matsukevich:2008}.
Characteristic features of these experiments were short laser
pulses exciting the atom and subsequent spontaneous emission of a
single photon. The second setting made use of an optical cavity to
efficiently direct photon emission into a predefined spatial mode
\cite{wilk:2007b}. Here, a vacuum-stimulated Raman adiabatic
passage technique was employed, with the driving laser pulse
controlling the photon shape \cite{kuhn:2002, mckeever:2004,
keller:2004}. Further research towards the deterministic entanglement of remote atoms, the teleportation of atomic qubit states, and the demonstration of quantum repeaters \cite{kimble:2008} would benefit if the advantages
of both settings could be combined in one setup, with
bandwidth-limited indistinguishable photons \cite{maunz:2007}
emitted into a well-defined spatio-temporal mode with high efficiency \cite{wilk:2007}.

In this letter, we report on the generation of single photons via short-pulse laser excitation of an atom coupled to an optical cavity.  The most intriguing feature of our excitation scheme is that the wave packet of the emitted photon is governed by the spectrum of the coupled atom-cavity system alone, independent of the excitation pulse shape and frequency.  In contrast
to free-space emission \cite{darquie:2005}, the coupled atom-cavity system
evolves with a coherent oscillatory energy exchange between the
atom and the cavity, damped by atomic and cavity decay. We record
the shape of the emitted single photon and investigate its
dependence from the detuning of the cavity with respect to the
atom. The observed oscillatory behavior is in excellent agreement
with theory and illustrates the fundamentals of cavity quantum electrodynamics at the single particle level.  Our technique further opens up new perspectives for shaping
single-photon wave packets \cite{difidio:2008}.

\begin{figure}
\includegraphics[width=1.0\columnwidth,keepaspectratio]{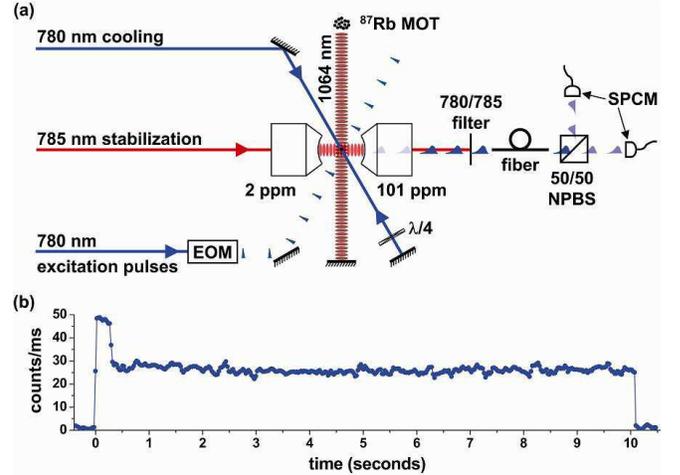}
\caption{\label{fig:setup} (a) $^{87}$Rb atoms are trapped within
the TEM$_{00}$ mode of the cavity at the intersection of a
standing wave dipole trap ($\lambda=1064$~nm) and an intracavity
dipole trap ($\lambda=785$~nm, also used for stabilizing the
cavity length). Two resonant beams at $\pm 45^{\circ}$ to the standing
wave trap and perpendicular to the cavity axis provide cooling and
fast pulse excitation. The cavity output is coupled to an optical
fiber and guided to the detection setup. SPCM: single photon
counting module, NPBS: non-polarizing beam splitter, $\lambda$/4:
quarter-wave plate, EOM: electro-optic modulator, MOT:
magneto-optical trap. (b) Measured cavity output photon stream of
a single atom with constant laser cooling. The standing wave trap
is turned on at $t=0$ with two atoms trapped during the first
300~ms. }
\end{figure}

Our new apparatus (similar to that described in \cite{nussmann:2005b, hijlkema:2007}, see
Figure~\ref{fig:setup}) uses an optical cavity operating
in the intermediate coupling regime with
\mbox{$(g_{\text{max}},\kappa,\gamma)/2\pi=(5.0,2.7,3.0)$~MHz.}
Here, $g_{\text{max}}$ denotes the atom-cavity coupling constant
averaged over all Zeeman sublevels of the $^{87}$Rb
\mbox{5S$_{1/2}~F=2\leftrightarrow$~5P$_{3/2}~F'=3$} transition
for an atom at a field antinode, $\gamma$ is the atomic polarization decay rate, and $\kappa$ is the cavity field
decay rate. The cavity is
frequency-stabilized to this atomic transition by means of a
reference laser ($\lambda = 785$~nm), which is itself locked to a
frequency comb. The cavity mirrors, each with 5~cm radius of curvature, are separated by 495~$\mu$m
giving a TEM$_{00}$-mode waist of 30~$\mu$m and a finesse of 56000
(mirror transmissions 2~ppm and 101~ppm, total losses 10~ppm). The
cavity output mode is coupled into a single-mode optical fiber and
directed to a Hanbury Brown-Twiss photon detection setup consisting of a non-polarizing beamsplitter and two single photon counting modules.  The
detection efficiency for a single photon present inside the cavity
is $\approx0.34$, which includes the directionality of the cavity
output ($\approx0.9$), spectral separation from stabilization
light and mode matching into the fiber ($\approx0.85$), and the
efficiency of the detectors ($\approx0.45$).

Single $^{87}$Rb atoms are loaded into the cavity mode from a
magneto-optical trap (MOT) via a running-wave dipole trap beam
\mbox{($\lambda = 1064$~nm)} with a focus between the MOT and the
cavity \cite{nussmann:2005}. When the atoms reach the cavity, the
transfer beam is replaced by a standing-wave beam ($\lambda =
1064$~nm). This beam is focused at the cavity mode and provides
strong spatial confinement along its axis (waist $16$~$\mu$m,
power $2.5$~W, potential depth $3$~mK). Additionally, the atoms
are confined along the cavity axis by the 785~nm cavity reference
laser (trap depth $70$~$\mu$K). Once trapped in the intra-cavity
2D optical lattice, the atoms are exposed to a retro-reflected
cooling laser beam incident perpendicularly to the cavity axis.
The cooling beam is near resonant with the
$F=2\leftrightarrow$~$F'=3$ transition (see below) and uses a
lin$\bot$lin polarization configuration. Light resonant with the
$F=1\leftrightarrow F'=2$ transition co-propagates with the
cooling beam for optical pumping out of the $F=1$ ground state. A
cavity emission signal of a single atom trapped and cooled inside
the cavity is shown in Figure~\ref{fig:setup}(b).

Long trapping times are observed under cavity-enhanced cooling conditions over the range $-70$~MHz~$\leq \Delta_{\text{cool}}/2\pi \leq -2$~MHz while
keeping $(\Delta_{\text{cav}}-\Delta_{\text{cool}})/2\pi=+5$~MHz, where $\Delta_{\text{cool}}$ and $\Delta_{\text{cav}}$ are the detunings of cooling laser and cavity with respect to the unperturbed \mbox{$F=2\leftrightarrow F'=3$} atomic resonance.  However, most important for this experiment, we are able to achieve trapping times of several seconds even outside the previously studied cavity-cooling regime \cite{nussmann:2005b}.  This occurs, for example, with a fixed cooling laser frequency at $\Delta_{\text{cool}}/2\pi=-50$~MHz and by varying the cavity frequency from $-45$~MHz~$\leq \Delta_{\text{cav}}/2\pi \leq +100$~MHz.  This suggests that Sisyphus-like cooling \cite{murr:2006} is the dominant mechanism, with resulting atomic temperatures comparable to those measured in \cite{nussmann:2005b}.  The advantage is now that the cavity frequency is a free parameter while maintaining sufficiently long atom trapping times.

Fast excitation of the atom-cavity system is accomplished by switching off the cooling light
and periodically driving the \mbox{$F=2\leftrightarrow F'=3$}
transition with $\approx3$~ns long laser pulses (FWHM). These
pulses are created by amplitude modulation of continuous-wave
light using a fiber-coupled electro-optic modulator (EOM, Jenoptik
model AM780HF). Due to the finite on:off ratio of the EOM~\cite{footnote:eomattenuation}, we detune the center frequency of the
excitation pulses from the cavity resonance by $-30$~MHz to
suppress continuous-wave excitations~\cite{hennrich:2000}.
Nevertheless, the atom is still excited by the short pulse
(measured bandwidth $\sim200$~MHz). Following
the fast excitation, ideally one photon will be produced because
the laser pulse is much shorter than the atomic lifetime (26~ns) and the
on-resonance build-up time for a field inside the cavity
($\tau_{\text{field}}\approx \pi/2g_{\text{max}}=50$~ns). As the system is driven on a
cycling transition,
no repumping is required before the next excitation pulse. This
scheme allows a pulse repetition rate of up to 5~MHz, limited by
the duration of the emitted single photon wave packet ($\lesssim
200$~ns). For all measurements presented here, the
pulse repetition rate is 670~kHz.

\begin{figure}
\includegraphics[width=1.0\columnwidth,keepaspectratio]{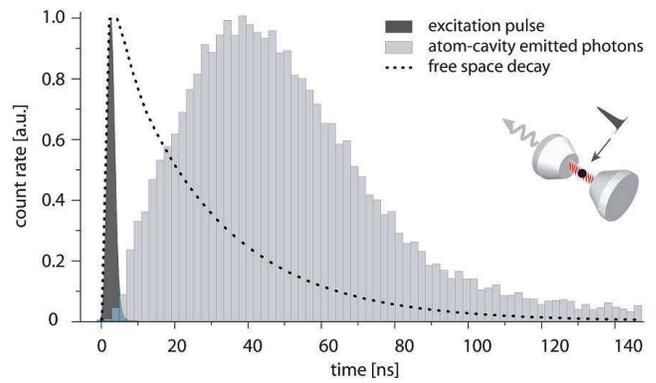}
\caption{\label{fig:pulses} Histogram: normalized
arrival time distribution of photons emitted from the trapped
atom-cavity system as a function of time after the excitation
pulse.  Solid filled curve: measured shape of the
excitation pulse. Dotted line: exponential decay
of the photon emission for an atom in free space.
Inset: Schematic of the atom-cavity system, excitation pulse, and
an emitted single photon.}
\end{figure}

Similar to the excitation of an atom in free space \cite{darquie:2005},
the fast laser pulse transfers the atom to the excited state
$\left|e\right\rangle$. However, the atom-cavity state
$\left|e,n=0\right\rangle$, where $n$ is the intracavity photon
number, is not an eigenstate of the coupled system. With atom and
cavity tuned into resonance, and for the moment neglecting dissipation, the system exhibits
oscillations according to
\begin{equation}
\left|\Psi(t)\right\rangle = \text{cos}\frac{\Omega t}{2}\left|e,0\right\rangle + \text{sin}\frac{\Omega t}{2}\left|g,1\right\rangle,
\label{eqn:oscs}
\end{equation}
where $\left|g\right\rangle$ is the atomic ground state and
$\Omega = 2g$ is the vacuum Rabi frequency \cite{book:haroche}.
This delays the peak of the photon emission compared to an atom in free space, as displayed 
in Fig.~\ref{fig:pulses}. For longer times, the oscillation is damped out due to atomic and
cavity decay resulting in a smooth wave packet envelope.

The single-photon nature of this excitation scheme is verified by
a measurement of the intensity correlation function
$g^{(2)}(\tau)$ evaluated from photons arriving within 200~ns
after the excitation pulse.  We observe a high suppression
($90\%$) of coincidence events at time $\tau=0$ demonstrating that
the protocol does indeed result in single photons. The remaining
coincident detections come from multiple trapped atoms ($\sim8\%$)
and dark counts of the photon detectors ($\sim2\%$). In contrast
to free-space emission \cite{darquie:2005}, the probability of
emitting two photons from the atom-cavity system from one excitation pulse is greatly
suppressed ($\sim10^{-4}$) as the single-photon field must first build up in the
cavity before the photon can be emitted.

Our measured probability of emitting a single photon into the
cavity mode following an excitation pulse is $\sim8\%$ for
$\Delta_\text{S}=\Delta_{\text{cav}}=80$~MHz and
$\Delta_\text{{pulse}}=50$~MHz. Here $\Delta_\text{S}$ denotes the
atomic AC-Stark shift due to the dipole trap, and
$\Delta_\text{{pulse}}$ is the detuning of the center frequency of
the excitation pulses (with respect to the unperturbed
\mbox{$F=2\leftrightarrow F'=3$} atomic transition). 
The photon production efficiency is limited 
by a number of technical difficulties.  For instance, as the atom is not prepared in a well defined initial Zeeman sublevel before each pulse, the average excitation probability saturates at $\sim50\%$ due to the indeterminate transition strengths.  Also, the average photon decay probability into free space ($\sim80\%$) is
a limitation since the atom-cavity cooperativity decreases due
to motion-induced variation of $g$. However, with appropriate choice
and improvement of excitation and cavity parameters, the photon emission
efficiency can in principle approach unity \cite{darquie:2005, kimble:2008}.

\begin{figure}
\includegraphics[width=1.0\columnwidth,keepaspectratio]{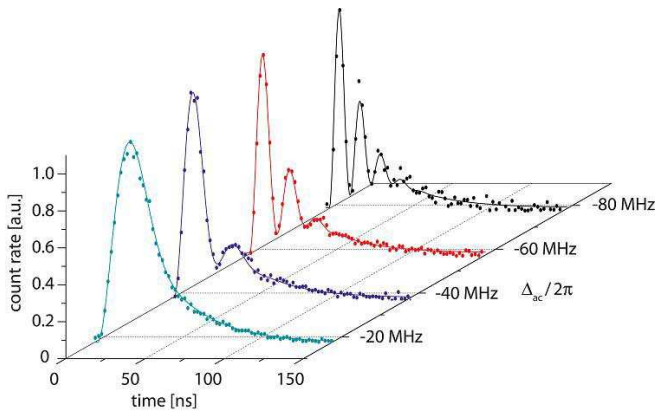}
\caption{\label{fig:osc} Measured arrival time distribution (dots)
of photons emitted from the cavity for
different atom-cavity detunings $\Delta_{\text{ac}}$.  The solid
lines are numerical fits (see text).}
\end{figure}

\begin{figure}
\includegraphics[width=1.0\columnwidth,keepaspectratio]{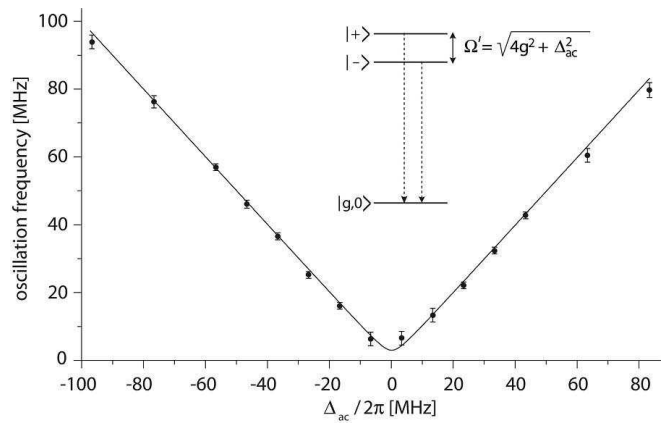}
\caption{\label{fig:hyp}  Measured oscillation frequency of the
emitted photons (dots) as a function of $\Delta_{\text{ac}}$. The
solid line shows the generalized Rabi frequency as a function of
detuning from Eq.~(\ref{eqn:hyperbola}).  Inset: Normal modes of
the coupled atom-cavity system. }
\end{figure}

The shape of the emitted photon can be further controlled by applying a detuning
between the Stark-shifted atomic resonance and the cavity resonance
$\Delta_{\text{ac}}=\Delta_{\text{S}}-\Delta_{\text{cav}}$
\cite{norris:1994}. The coupled system then oscillates between the
states $\left|e,0\right\rangle$ and $\left|g,1\right\rangle$ at a
frequency
\begin{equation}
\Omega' = \sqrt{4g^2+\Delta_{\text{ac}}^2}. \label{eqn:hyperbola}
\end{equation}
Note that because $\gamma\approx\kappa$, shifts of the oscillation
frequency due to damping are negligible \cite{book:haroche}.
We observe these oscillations for a single atom trapped within a standing wave, however, the position-dependent Stark shift reduces the measured contrast.  A more constant, and in particular smaller, Stark shift is maintained in a running-wave trap
($\Delta_{\text{S}}\approx2\pi\times17$~MHz).  
We ensure that less than one atom is present in the cavity at any time by again measuring the $g^{(2)}(\tau)$
correlation function. As seen in Fig.~\ref{fig:osc}, the emitted photon
wave packets exhibit modulations according to the population
dynamics of state $\left|g,1\right\rangle$. Note that no
externally applied driving field is present during the
oscillations, and the measured features are not due to many-photon
or many-atom effects \cite{brecha:1995, brune:1996, mielke:1997}.

We find good agreement between the measured photon wave packet
shapes and an analytical model (solid lines in Fig.
\ref{fig:osc}). The model consists of an oscillating term for the
occupation of state $\left|g,1\right\rangle$ and an exponential
damping term due to atomic and cavity decay. Further, we include
the experimental shot-to-shot uncertainty of $\Omega'$ using a
Gaussian distribution of fixed width (10~MHz), which accounts for
position-dependent Stark shifts and variations of $g$. We
numerically fit this model to the data and extract values of
$\Omega'$ as a function of atom-cavity detuning
(Fig.~\ref{fig:hyp}). The extracted frequencies match well the
predicted hyperbolic function (Eq.~\ref{eqn:hyperbola}) with a reduced chi-squared $\chi^2=1.13$.
These measurements can also be understood as time-domain
normal-mode spectroscopy of a coupled single-atom-cavity system in
the optical regime. In this interpretation, the observed
modulation originates from a quantum beat at the frequency
difference between the energy levels of the first pair of
atom-cavity dressed states (Fig.~\ref{fig:hyp} inset) \cite{haroche:1973}.

\begin{figure}
\includegraphics[width=1.0\columnwidth,keepaspectratio]{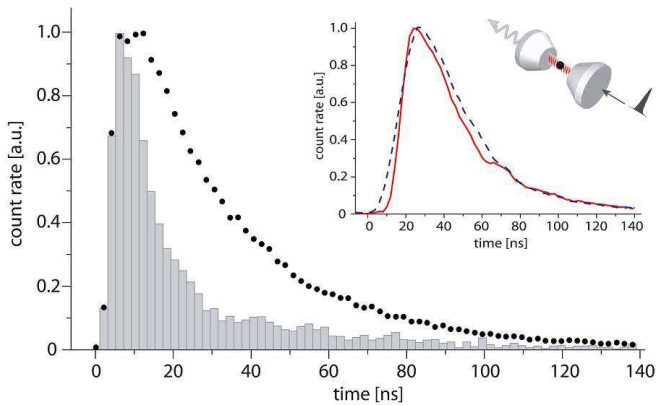}
\caption{\label{fig:pumpcavity} Measured photon arrival time
distribution for excitation of the cavity with no atoms (filled
dots) and $\lesssim1$ atom (histogram).  The inset compares the
temporal evolution of state $\left|g,1\right\rangle$ for
transverse pumping (solid line) with that of
$\left|e,0\right\rangle$ for cavity pumping (dashed line),
illustrating the complementary dynamics.}
\end{figure}

A complementary illustration of the coherent oscillations between
the atom and the cavity is investigated by pumping the
system along the axis of the cavity (Fig.~\ref{fig:pumpcavity}).
In this case, a weak 3~ns laser pulse excites the atom-cavity system such that when an output photon is detected, we know the system was in $\left|g,1\right\rangle$ immediately after excitation. We observe a
deviation from a pure exponential decay of the intra-cavity field
as the energy is temporarily stored in the atom. Quantitatively,
we can retrieve the temporal evolution of the population in state
$\left|e,0\right\rangle$ by subtracting this signal from the
exponential decay of the empty cavity at rate 2$\kappa$. The inset
of Fig.~\ref{fig:pumpcavity} compares this difference signal with
the measured evolution of $\left|g,1\right\rangle$ from the
experiment with transverse excitation. The nearly identical time
dependence of the two signals testifies to the coherent
exchange of energy between atom and field.

In future experiments, by terminating the atom-cavity oscillations with a suitably timed atomic de-excitation pulse, the fast excitation technique should allow one to design single photons with duration shorter than the system's decay time \cite{difidio:2008}. 
This includes the possibility to generate time-symmetric photons important 
for quantum networking \cite{cirac:1997}.
Additionally, a single photon in a superposition
of two tunable frequencies, as demonstrated here, may be useful as a
frequency qubit \cite{duan:2006, lan:2007}. Our scheme may also find
application in the investigation of higher-lying dressed states in
cavity QED systems \cite{schuster:2008} and for the deterministic generation of multi-photon Fock states \cite{kubanek:2008}. Finally, our fast excitation technique can improve existing atom-photon entanglement experiments \cite{wilk:2007b} by reducing unwanted
multiple-photon events, and can be extended to multi-photon entanglement protocols \cite{schon:2005}.

\begin{acknowledgments}
The authors thank K. Murr for useful discussions and B. Bernhard,
T. Wilken and R. Holzwarth for providing the frequency comb signal.
This work was partially supported by the Deutsche
Forschungsgemeinschaft (Research Unit 635) and the European Union
(IST program, SCALA). G. L.-K. acknowledges support from the Rosa
Luxemburg Foundation. D. L. M. acknowledges support from the
Alexander von Humboldt Foundation.
\end{acknowledgments}


\begin{thebibliography}{10}

\bibitem{zoller:2005}
P.~Zoller, {\em et~al\/}.
\newblock Eur. Phys. J. D {\bf 36}, 203 (2005).

\bibitem{blinov:2004}
B.~B. Blinov, D.~L. Moehring, L.-M. Duan, and C.~Monroe.
\newblock Nature {\bf 428}, 153 (2004).

\bibitem{volz:2006}
J.~Volz, {\em et~al\/}.
\newblock Phys. Rev. Lett. {\bf 96}, 030404 (2006).

\bibitem{moehring:2007b}
D.~L. Moehring, {\em et~al\/}.
\newblock Nature {\bf 449}, 68 (2007).

\bibitem{matsukevich:2008}
D.~N. Matsukevich, {\em et~al\/}.
\newblock Phys. Rev. Lett. {\bf 100}, 150404 (2008).

\bibitem{wilk:2007b}
T.~Wilk, S.~C. Webster, A.~Kuhn, and G.~Rempe.
\newblock Science {\bf 317}, 488 (2007).

\bibitem{kuhn:2002}
A.~Kuhn, M.~Hennrich, and G.~Rempe.
\newblock Phys. Rev. Lett. {\bf 89}, 067901 (2002).

\bibitem{mckeever:2004}
J.~McKeever, {\em et~al\/}.
\newblock Science {\bf 303}, 1992 (2004).

\bibitem{keller:2004}
M.~Keller, {\em et~al\/}.
\newblock Nature {\bf 431}, 1075 (2004).

\bibitem{kimble:2008}
H.~J. Kimble.
\newblock Nature {\bf 453}, 1023 (2008).

\bibitem{maunz:2007}
P.~Maunz, {\em et~al\/}.
\newblock Nature Physics {\bf 3}, 538 (2007).

\bibitem{wilk:2007}
T.~Wilk, {\em et~al\/}.
\newblock Phys. Rev. Lett. {\bf 98}, 063601 (2007).

\bibitem{darquie:2005}
B.~Darqui{\'{e}}, {\em et~al\/}.
\newblock Science {\bf 309}, 454 (2005).

\bibitem{difidio:2008}
C.~DiFidio, W.~Vogel, M.~Khanbekyan, and D.-G. Welsch.
\newblock Phys. Rev. A {\bf 77}, 043822 (2008).

\bibitem{nussmann:2005b}
S.~Nu{\ss}mann, {\em et~al\/}.
\newblock Nature Physics {\bf 1}, 122 (2005).

\bibitem{hijlkema:2007}
M.~Hijlkema, {\em et~al\/}.
\newblock Nature Physics {\bf 3}, 253 (2007).

\bibitem{nussmann:2005}
S.~Nu{\ss}mann, {\em et~al\/}.
\newblock Phys. Rev. Lett. {\bf 95}, 173602 (2005).

\bibitem{murr:2006}
K.~Murr, {\em et~al\/}.
\newblock Phys. Rev. A {\bf 73}, 063415 (2006).

\bibitem{footnote:eomattenuation}
The on:off ratio is measured to be greater than 1000:1. This value is reached
  within 8~ns of the peak power.

\bibitem{hennrich:2000}
M.~Hennrich, T.~Legero, A.~Kuhn, and G.~Rempe.
\newblock Phys. Rev. Lett. {\bf 85}, 4872 (2000).

\bibitem{book:haroche}
{S. Haroche} and {J. M. Raimond}.
\newblock {\em Exploring the Quantum -- Atoms, Cavities and Photons\/} (Oxford
  University Press, Oxford) (2006).

\bibitem{norris:1994}
T.~B. Norris, {\em et~al\/}.
\newblock Phys. Rev. B {\bf 50}, 14663 (1994).

\bibitem{brecha:1995}
R.~J. Brecha, {\em et~al\/}.
\newblock J. Opt. Soc. Am. B {\bf 12}, 2329 (1995).

\bibitem{brune:1996}
M.~Brune, {\em et~al\/}.
\newblock Phys. Rev. Lett. {\bf 76}, 1800 (1996).

\bibitem{mielke:1997}
S.~L. Mielke, G.~T. Foster, J.~Gripp, and L.~A. Orozco.
\newblock Opt. Lett. {\bf 22}, 325 (1997).

\bibitem{haroche:1973}
S.~Haroche, J.~A. Paisner, and A.~L. Schawlow.
\newblock Phys. Rev. Lett. {\bf 30}, 948 (1973).

\bibitem{cirac:1997}
J.~I. Cirac, P.~Zoller, H.~J. Kimble, and H.~Mabuchi.
\newblock Phys. Rev. Lett. {\bf 78}, 3221 (1997).

\bibitem{duan:2006}
L.-M. Duan, {\em et~al\/}.
\newblock Phys. Rev. A {\bf 73}, 062324 (2006).

\bibitem{lan:2007}
S.-Y. Lan, {\em et~al\/}.
\newblock Phys. Rev. Lett. {\bf 98}, 123602 (2007).

\bibitem{schuster:2008}
I.~Schuster, {\em et~al\/}.
\newblock Nature Physics {\bf 4}, 382 (2008).

\bibitem{kubanek:2008}
A.~Kubanek, {\em et~al\/}.
\newblock Phys. Rev. Lett. (in press) (2008).

\bibitem{schon:2005}
C.~Sch{\"{o}}n, {\em et~al\/}.
\newblock Phys. Rev. Lett. {\bf 95}, 110503 (2005).

\end{thebibliography}
\end{document}